\documentclass[sigconf,nonacm]{acmart}

\usepackage{subfigure}
\usepackage{siunitx}
\usepackage{enumitem}

\usepackage{algorithm}
\usepackage[noend]{algpseudocode}
\let\ReturnInline\Return
\renewcommand{\Return}{\State\ReturnInline}
\algrenewcommand\algorithmicrequire{$\rhd$}
\algrenewcommand\algorithmicensure{$\square$}

\AtBeginDocument{%
  \providecommand\BibTeX{{%
    \normalfont B\kern-0.5em{\scshape i\kern-0.25em b}\kern-0.8em\TeX}}}

\setcopyright{acmcopyright}
\copyrightyear{2018}
\acmYear{2018}
\acmDOI{XXXXXXX.XXXXXXX}

\acmConference[Conference acronym 'XX]{Make sure to enter the correct
  conference title from your rights confirmation emai}{June 03--05,
  2018}{Woodstock, NY}
\newcommand{\ignore}[1]{}

\begin{document}

%% Full title of the paper.
\title[GVE-LPA: Fast Label Propagation Algorithm (LPA) for Community Detection in Shared Memory Setting]{GVE-LPA: Fast Label Propagation Algorithm (LPA) for Community Detection in Shared Memory Setting}

%% Short title to be used in page headers (optional).
% \title[short title]{full title}
% \subtitle{Something other than the title}

%% Authors and their affiliations.
\author{Subhajit Sahu}
\email{subhajit.sahu@research.iiit.ac.in}
\affiliation{%
  \institution{IIIT Hyderabad}
  \streetaddress{Professor CR Rao Rd, Gachibowli}
  \city{Hyderabad}
  \state{Telangana}
  \country{India}
  \postcode{500032}
}

%% Concise author list in page headers.
%\renewcommand{\shortauthors}{Sahu, Kothapalli, and Banerjee, et al.}

%% Show page numbers.
\settopmatter{printfolios=true}

%% Short summary of the work to be presented in the article.
\begin{abstract}
Community detection is the problem of identifying natural divisions in networks. Efficient parallel algorithms for this purpose are crucial in various applications, particularly as datasets grow to substantial scales. This technical report presents an optimized parallel implementation of the Label Propagation Algorithm (LPA), a high speed community detection method, for shared memory multicore systems. On a server equipped with dual 16-core Intel Xeon Gold 6226R processors, our LPA, which we term as GVE-LPA, outperforms FLPA, igraph LPA, and NetworKit LPA by $139\times$, $97000\times$, and $40\times$ respectively - achieving a processing rate of $1.4 B$ edges/s on a $3.8 B$ edge graph. In addition, GVE-LPA scales at a rate of $1.7\times$ every doubling of threads.
\end{abstract}


% \ccsdesc[500]{Theory of computation~Parallel algorithms}
% \ccsdesc[500]{Theory of computation~Graph algorithms analysis}

%% Pick words that accurately describe the work being presented.


%% Process the author and title information.
\maketitle

\section{Introduction}
\label{sec:introduction}
Community detection is the problem of identifying cohesive groups of vertices, also known as clusters, in complex networks that minimize cuts and maximize internal density \cite{garza2019community}.\ignore{This problem finds applications in various domains, including drug discovery, protein annotation, topic exploration, anomaly detection, and criminal identification.} Communities identified are intrinsic when based on network topology alone, and are disjoint when each vertex belongs to only one community \cite{com-gregory10}. One of the difficulties in community detection is the lack of apriori knowledge on the number and size distribution of communities \cite{com-blondel08}.

Community detection finds applications in various domains. In drug discovery, it helps identify groups of similar compounds or target proteins, thereby facilitating the discovery of new therapeutic agents \cite{ma2019comparative, udrescu2020uncovering}. In the health domain, it aids in understanding the dynamics of groups susceptible to epidemic diseases \cite{salathe2010dynamics}, detecting diseases like lung cancer \cite{bechtel2005lung}, and categorizing tumor types using genomic datasets \cite{haq2016community}. Additionally, it has been applied to elucidate the structure and evolution of metabolic networks \cite{pfeiffer2005evolution, kim2009centralized}, Gene Regulatory Networks (GRNs) \cite{rivera2010nemo}, and Lateral Gene Transfer (LGT) networks \cite{popa2011directed}. Another significant application of community detection is in the analysis of human brain networks \cite{bullmore2009complex, he2010graph}. In ecological studies, it is used to determine if food webs are organized into compartments, where species within the same compartment frequently interact among themselves but have fewer interactions with species in different compartments \cite{krause2003compartments, rezende2009compartments, guimera2010origin}.

The \textit{Label Propagation Algorithm (LPA)}, also known as RAK, \cite{com-raghavan07} is a popular diffusion-based approach for identifying communities. It is faster and more scalable than the Louvain method \cite{com-blondel08}, another high-quality community detection algorithm, as it does not require repeated optimization steps and is easier to parallelize \cite{com-newman04, com-raghavan07}. Due to this, LPA has been used in a number of other graph problems and applications, such as, finding connected components \cite{stergiou2018shortcutting}, graph partitioning \cite{slota2014pulp, wang2014partition, meyerhenke2014partitioning, meyerhenke2016partitioning, meyerhenke2017parallel, bae2020label, akhremtsev2020high, slota2020scalable, zhang2020multilevel}, hypergraph partitioning \cite{henne2015label, gottesburen2021scalable}, graph coarsening \cite{valejo2020coarsening}, vertex reordering and graph compression \cite{boldi2011layered}, unsupervised part-of-speech tagging \cite{das2011unsupervised}, sectionalizing power systems \cite{aziz2023novel}. Further, Peng et al. \cite{peng2014accelerating} observe that, while LPA achieves lower modularity scores (a metric for assessing community quality \cite{com-newman04}) than the Louvain method --- a lower modularity score for LPA implies that it may struggle to properly cluster nodes that are not well-connected to any particular community--- it attains the highest Normalized Mutual Information (NMI) score when compared to the ground truth.

A large number of LPA variants have also been proposed \cite{li2015parallel, farnadi2015scalable, li2015detecting, shen2016topic, mohan2017scalable, zhang2017label, berahmand2018lp, ma2018psplpa, sattari2018spreading, zheng2018improved, xu2019distributed, zarei2020detecting, maleki2020dhlp, zhang2020lilpa, el2021wlni, roghani2021pldls, zhang2023large}. Existing studies on LPA propose a number of algorithmic optimizations and parallelization techniques, but there continue to be a number of issue with each implementation. Further, the race for AI development has caused the prices of GPUs to skyrocket. These issues, and the large applicability of LPA, motivates us to present GVE-LPA,\footnote{\url{https://github.com/puzzlef/rak-communities-openmp}} an efficient implementation of LPA for multicore architectures.

\section{Related work}
\label{sec:related}
For parallelization of LPA, vertex assignment has been achieved with guided scheduling \cite{staudt2015engineering}, parallel bitonic sort \cite{soman2011fast}, and pre-partitioning of the graph \cite{kuzmin2015parallelizing}. Additional improvements upon the LPA include using a stable (non-random) mechanism of label choosing in the case of multiple best labels \cite{com-xing14}, addressing the issue of monster communities \cite{com-berahmand18, com-sattari18}, identifying central nodes and combining communities for improved modularity \cite{com-you20}, and using frontiers with alternating push-pull to reduce the number of edges visited and improve solution quality \cite{com-liu20}.\ignore{A number of variants of LPA have been proposed, but the original formulation is still the simplest and most efficient \cite{garza2019community}.}

A few open source implementations\ignore{and software packages} have been developed for community detection using LPA. Fast Label Propagation Algorithm (FLPA) \cite{traag2023large} is a fast variant of the LPA, which utilizes a queue-based approach to process only vertices with recently updated neighborhoods. NetworKit \cite{staudt2016networkit} is a software package designed for analyzing the structural aspects of graph data sets with billions of connections. It is implemented as a hybrid with C++ kernels and a Python frontend, and includes parallel implementation of LPA. igraph \cite{csardi2006igraph} is a similar package, written in C, with Python, R, and Mathematica frontends. It is widely used in\ignore{academic} research, and includes an implementation of LPA.

We now discuss the implementation of LPA in NetworKit, igraph, and FLPA. The parallel implementation of LPA in NetworKit is given in \texttt{NetworKit::PLP::run()}. This function starts with a unique label for each node. To check for convergence, they use a tolerance of $10^{-5}$, i.e., the algorithm converges once the labels of less than $0.001\%$ of vertices change. This is also referred to as threshold heuristic. To track active nodes, a boolean flag vector is employed, serving as a vertex pruning optimization. A parallel for loop, which uses OpenMP's \textit{guided} schedule, then processes only the active nodes. For storing label weights, an \texttt{std::map} is used for each vertex. In contrast, the LPA implementation of igraph and FLPA is sequential. igraph's LPA implementation, found in the function \texttt{igraph\_community\_label\_propagation()}, shuffles the node order in each iteration. It keeps track of the dominant labels for each new neighbor of a vertex and selects one of the dominant labels using a random number generator. The hashtable is cleared by iterating through the neighbors. Unlike NetworKit, igraph does not use the pruning optimization. It alternates between label updating and control iterations, checking if the current label of a node is not dominant during control iterations, with no tolerance for non-dominant labels. The FLPA implementation in igraph, detailed in the function \texttt{igraph\_i\_community\_label\_propagation()}, uses a dequeue for managing a set of unprocessed vertices, which indicates convergence when empty. FLPA does not shuffle for random node order. If a label change occurs, FLPA considers neighbors that are not in the same community.

However, we identify a few issues with the LPA implementations of NetworKit, igraph, and FLPA. In order to assign a unique label to each node in the graph, NetworKit uses OpenMP's plain parallel for, i.e., it uses \textit{static} loop scheduling with a chunk size of $1$. This can result in false sharing as threads make writes to consecutive locations in memory. Next, to keep track of the weights associated with each label for a given vertex, they use an \texttt{std::map}. Our experiments show that this is quite inefficient. We, in contrast, use a per-thread keys list and a per-thread full-size values array, as our hashtable. In order to check for convergence, NetworKit uses a tolerance of $10^{-5}$. However, we observe that a tolerance of $10^{-2}$ is generally obtains community of nearly the same quality (in terms of modularity), but converges much faster. Further, NetworKit uses atomic operations on a shared variable to keep a count of the number of updated vertices. This can introduce contention. We, instead, make use of parallel reduce for this. Finally, NetworKit uses a boolean\ignore{flag} vector to keep track of the set of active nodes. However, our experiments show that using an 8-bit integer flag vector is more efficient\ignore{, despite its larger memory footprint}.

We now discuss issues in the LPA implementation of igraph. igraph randomly shuffles the processing order of nodes in each iteration. This can get quite expensive. igraph checks for convergence in alternate iterations, and algorithm is considered to have converged only if the labels of all nodes are dominant, i.e., are the most popular. Thus it can take a large number of iterations to converge, with minimal gain in community quality. In addition, igraph does not use vertex pruning optimization, i.e., it does not maintain the set of active nodes. Finally, given multiple dominant labels, igraph select a random dominant label as the label of the given node --- however, random number generation is slow. We\ignore{instead} use the dominant label with the smallest ID as the label of the given vertex.

Finally, we discuss the issues in FLPA's implementation. Similar to igraph LPA, given multiple dominant labels, FLPA selects a random dominant label as the label of the given node --- again, random number generation is slow. Further, FLPA considers the algorithm to have converged only when there are no active nodes. Thus it can take a large number of iterations to converge, with minimal gain in community quality.

\section{Preliminaries}
\label{sec:preliminaries}
Consider an undirected graph $G(V, E, w)$, where $V$ is the set of vertices, $E$ is the set of edges, and $w_{ij} = w_{ji}$ represents the weight associated with each edge. For unweighted graphs, we assume a unit weight for each edge, i.e., $w_{ij} = 1$. The neighbors of a vertex $i$ are denoted as $J_i = {j\ |\ (i, j) \in E}$, the weighted degree of each vertex as $K_i = \sum_{j \in J_i} w_{ij}$, the total number of vertices as $N = |V|$, the total number of edges as $M = |E|$, and the sum of edge weights in the undirected graph as $m = \sum_{i, j \in V} w_{ij}/2$.

\subsection{Community detection}

Disjoint community detection involves the identification of a community membership mapping, $C: V \rightarrow \Gamma$, where each vertex $i \in V$ is assigned a community ID $c \in \Gamma$, with $\Gamma$ representing the set of community IDs. We denote the vertices belonging to a community $c \in \Gamma$ as $V_c$, and the community to which a vertex $i$ belongs as $C_i$. Additionally, we denote the neighbors of vertex $i$ within community $c$ as $J_{i \rightarrow c} = {j\ |\ j \in J_i\ and\ C_j = c}$, the sum of edge weights for those neighbors as $K_{i \rightarrow c} = \sum_{j \in J_{i \rightarrow c}} w_{ij}$, the edge weight within a community $c$ as $\sigma_c = \sum_{(i, j) \in E\ and\ C_i = C_j = c} w_{ij}$, and the total edge weight of a community $c$ as $\Sigma_c = \sum_{(i, j) \in E\ and\ C_i = c} w_{ij}$ \cite{com-leskovec21}.

\subsection{Modularity}

Modularity functions as a metric for evaluating the quality of communities identified by heuristic-based community detection algorithms. It is calculated as the difference between the fraction of edges within communities and the expected fraction under random edge distribution. It lies in a range of $[-0.5, 1]$, where higher values signify superior quality \cite{com-brandes07}.\ignore{The optimization of this metric theoretically leads to the optimal grouping \cite{com-newman04, com-traag11}.} The modularity $Q$ for the identified communities is computed using Equation \ref{eq:modularity}, where $\delta$ denotes the Kronecker delta function ($\delta (x,y)=1$ if $x=y$, $0$ otherwise).\ignore{The \textit{delta modularity} of moving a vertex $i$ from community $d$ to community $c$, denoted as $\Delta Q_{i: d \rightarrow c}$, can be computed using Equation \ref{eq:delta-modularity}.}

\begin{equation}
\label{eq:modularity}
  Q
  = \frac{1}{2m} \sum_{(i, j) \in E} \left[w_{ij} - \frac{K_i K_j}{2m}\right] \delta(C_i, C_j)
  = \sum_{c \in \Gamma} \left[\frac{\sigma_c}{2m} - \left(\frac{\Sigma_c}{2m}\right)^2\right]
\end{equation}

\ignore{\begin{equation}
\label{eq:delta-modularity}
  \Delta Q_{i: d \rightarrow c}
  = \frac{1}{m} (K_{i \rightarrow c} - K_{i \rightarrow d}) - \frac{K_i}{2m^2} (K_i + \Sigma_c - \Sigma_d)
\end{equation}}

\subsection{Label Progagation Algorithm (LPA)}
\label{sec:about-rak}

LPA \cite{com-raghavan07} is a popular diffusion-based method for identifying communities of medium quality in large networks. It is simpler, faster, and more scalable compared to the Louvain method \cite{com-blondel08}. In LPA, each vertex $i$ is starts with a unique label (community id) $C_i$. During each iteration, vertices adopt the label with the highest interconnecting weight, as shown in Equation \ref{eq:lpa}. Through this iterative process, a consensus among densely connected groups of vertices is achieved\ignore{, forming communities}. The algorithm converges when at least $1-\tau$ fraction of vertices (where $\tau$ is the tolerance parameter) maintain their community membership. LPA exhibits a time complexity of $O(L |E|)$ and a space complexity of $O(|V| + |E|)$, with $L$ denoting the number of iterations performed \cite{com-raghavan07}. We have experimented with COPRA \cite{com-gregory10}, SLPA \cite{com-xie11}, and LabelRank \cite{com-xie13}, but found LPA to be the most performant, while yielding communities of equivalent quality \cite{sahu2023selecting}.

\begin{equation}
\label{eq:lpa}
  C_i =\ \underset{c\ \in \ \Gamma}{\arg\max} { \sum_{j \in J_i\ |\ C_j = c} w_{ij} }
\end{equation}

\section{Approach}
\label{sec:approach}
\subsection{Optimizations for LPA}
\label{sec:lpa}

We use an \textit{asynchronous} parallel implementation of LPA, wherein threads operate independently on distinct regions of the graph. This facilitates quicker convergence, but may introduce greater variability in the final result. Further, we observe that parallel LPA obtains communities of higher quality than its sequential implementation, possibly due to randomization. We allocate a separate hashtable per thread to keep track of the total weight of each unique label linked to a vertex.

\begin{figure*}[hbtp]
  \centering
  \subfigure{
    \label{fig:rak-opt--all}
    \includegraphics[width=0.98\linewidth]{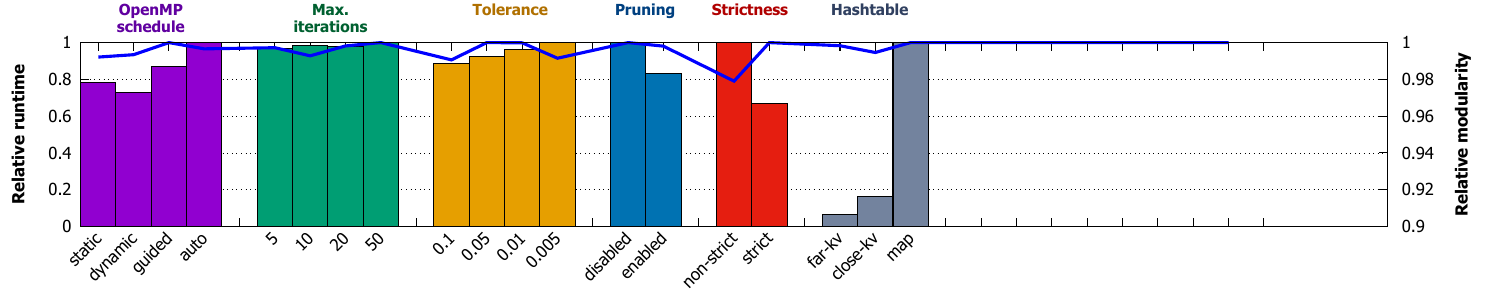}
  } \\[-2ex]
  \caption{Impact of\ignore{various} parameter controls and optimizations on the runtime and result quality (modularity) of LPA. We show the impact of each optimization upon the relative runtime on the left Y-axis, and upon the relative modularity on the right Y-axis.}
  \label{fig:rak-opt}
\end{figure*}

We now evaluate alternatives for a few optimizations, on the system mentioned in Section \ref{sec:setup}, for every graph in the dataset (Table \ref{tab:dataset}) --- conducting them five times on each graph to minimize the influence of noise. We then calculate the geometric mean for the runtime and arithmetic mean for the modularity, and represent them as ratios within each optimization category, as shown in Figure \ref{fig:rak-opt}. While none of our optimizations are particularly novel, we carefully address each one and combine them to achieve superior performance compared to other implementations of LPA.\ignore{We believe our implementation will aid the community in advancing more rapidly and tackling more challenges.}

\ignore{Our optimizations include using OpenMP's \verb|dynamic| loop schedule, setting an initial tolerance of $0.05$, enabling vertex pruning, employing the strict version of LPA, and using fast collision-free per-thread hashtables which are well separated in their memory addresses (\textit{Far-KV}). See below for the details on each optimization.}

\subsubsection{Adjusting OpenMP loop schedule}

We attempt\ignore{\textit{static}, \textit{dynamic}, \textit{guided}, and \textit{auto}} various loop scheduling approaches of OpenMP --- each using a chunk size of $2048$ (we observe this strikes a balance between minimizing load imbalance and reducing scheduling overhead) to parallelize LPA. We consider OpenMP's \verb|dynamic| loop schedule to be the best choice, due to its ability of work balancing among threads, and because it yields a runtime reduction of $27\%$ when compared to OpenMP's \textit{auto} loop schedule, while incurring only a $0.7\%$ reduction in the modularity of obtained communities (likely to be just noise).
% This chunk size, determined through experimentation, strikes an optimal balance between minimizing load imbalance and reducing scheduling overhead in the parallelization of LPA. 

\subsubsection{Limiting the number of iterations}

Restricting the number of iterations of LPA can ensure its termination within a reasonable number of iterations, but choosing a small limit may worsen the quality of communities obtained. Our results suggest that limiting the maximum number of iterations to $20$ strikes a good balance between runtime and modularity.

\subsubsection{Adjusting tolerance}

Using a small tolerance allows the algorithm to explore broader possibilities for community assignments, but comes at the cost of increased runtime. We find an initial tolerance of $0.05$ to be suitable. A tolerance of $0.1$ may also be acceptable, but provides a very small gain in performance when compared to a tolerance of $0.05$.

\subsubsection{Vertex pruning}

\begin{figure}[hbtp]
  \centering
  \subfigure{
    \label{fig:rak-pruning--all}
    \includegraphics[width=0.78\linewidth]{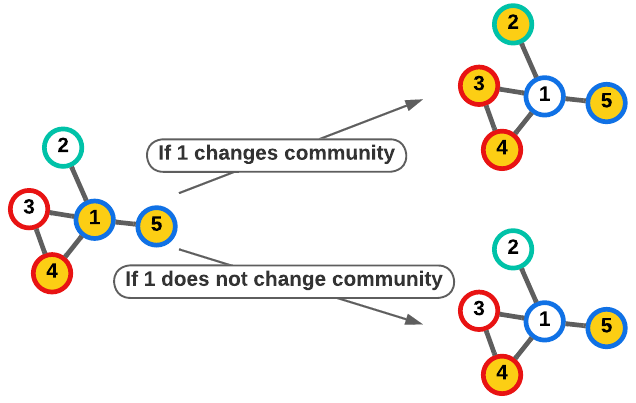}
  } \\[-2ex]
  \caption{Illustration of vertex pruning optimization: Once vertex $1$ is processed, it is unmarked. If vertex $1$ changes its community, its neighbors are marked for processing. Community membership of each vertex is indicated by border color, and marked vertices are highlighted in yellow \cite{sahu2023gvelouvain}.}
  \label{fig:rak-pruning}
\end{figure}

Vertex pruning is a method utilized to minimize unnecessary computation. In this approach, when a vertex alters its community, it assigns its neighbors for processing. Once a vertex has been processed, it is labeled as ineligible for further processing. However, this procedure incurs an additional overhead due to the marking and unmarking of vertices. Based on our findings, the employment of vertex pruning justifies this overhead and results in a performance enhancement of $17\%$. An illustration of vertex pruning optimization is shown in Figure \ref{fig:rak-pruning}.

\subsubsection{Picking the best label}

When there exist multiple labels connected to a vertex with maximum weight, we may randomly pick one of them (non-strict LPA), or pick only the first of them (strict LPA). We implement non-strict LPA using a simple modulo operator on the label id, as we observe that using \textit{xorshift} based random number generator does not provide any advantage. Results indicate that the strict version of LPA is $1.5\times$ faster than the non-strict approach, while also offering a gain in modularity of $2.1\%$.

\subsubsection{Hashtable design}

\begin{figure}[hbtp]
  \centering
  \subfigure{
    \label{fig:rak-hashtable--all}
    \includegraphics[width=0.88\linewidth]{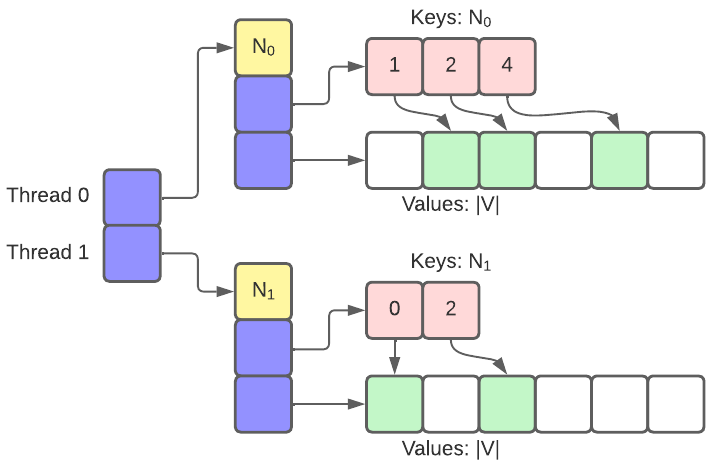}
  } \\[-2ex]
  \caption{Illustration of collision-free per-thread hashtables with well separated memory addresses (Far-KV) for two threads. Each hashtable comprises a keys vector, a values vector (of size $|V|$), and a key count ($N_0$/$N_1$). The value corresponding to each key is stored/accumulated at the index indicated by the key. As the key count of each hashtable is updated independently, we allocate it separately on the heap to avoid false cache sharing \cite{sahu2023gvelouvain}.}
  \label{fig:rak-hashtable}
\end{figure}

One can utilize C++'s inbuilt map as per-thread (independent) hashtables for the LPA algorithm. However, this exhibits poor performance. Therefore, we employ a key-list and a collision-free full-size values array to dramatically improve performance. However, if the memory addresses of the hashtables are nearby (\textit{Close-KV}), even if each thread uses its own hashtable exclusively, the performance is not as high. This is possibly due to false cache-sharing. Alternatively, if we ensure that the memory address of each hashtable is farther away (\textit{Far-KV}), the performance improves. Our results indicate that \textit{Far-KV} has the best performance and is $15.8\times$ times faster than \textit{Map}, and $2.6\times$ times faster than \textit{Close-KV} with LPA. An illustration of \textit{Far-KV} hashtable is in Figure \ref{fig:rak-hashtable}.

\subsection{Our optimized LPA implementation}

We now explain the implementation of GVE-LPA in Algorithm \ref{alg:rak}. Its main step is the \texttt{lpa()} function (lines \ref{alg:rak--main-begin}-\ref{alg:rak--main-end}), which takes the input graph $G$ and assigns community memberships (or labels) $C$ to each vertex. In lines \ref{alg:rak--init-begin}-\ref{alg:rak--init-end}, we first initialize the labels $C$ of each vertex in $G$, and mark all vertices as unprocessed. We then perform iterations to propagate labels based on the weighted influence of neighboring vertices, limited to $MAX\_ITERATIONS$ (lines \ref{alg:rak--iters-begin}-\ref{alg:rak--iters-end}). In each iteration, we invoke the \texttt{lpaMove()} function to perform label propagation, and count the number of nodes with updated labels $\Delta N$ (line \ref{alg:rak--propagate}). If the ratio of $\Delta N$ to the total number of nodes $N$ is within a specified tolerance $\tau$, convergence has been achieved, and we terminate the loop (line \ref{alg:rak--converged}). Upon completing all iterations, we return the final labels $C$ (line \ref{alg:rak--main-return}).

\begin{algorithm}[hbtp]
\caption{GVE-LPA: Our parallel LPA.}
\label{alg:rak}
\begin{algorithmic}[1]
\Require{$G$: Input graph}
\Require{$C$: Community membership of each vertex}
\Ensure{$H_t$: Collision-free per-thread hashtable}
\Ensure{$l_i$: Number of iterations performed}
\Ensure{$\tau$: Per iteration tolerance}

\Statex

\Function{lpa}{$G$} \label{alg:rak--main-begin}
  \State Vertex membership: $C \gets [0 .. |V|)$ \label{alg:rak--init-begin}
  \State Mark all vertices in $G$ as unprocessed \label{alg:rak--init-end}
  \ForAll{$l_i \in [0 .. \text{\small{MAX\_ITERATIONS}})$} \label{alg:rak--iters-begin}
    \State $\Delta N \gets lpaMove(G, C)$ \label{alg:rak--propagate}
    \If{$\Delta N/N \le \tau$} \textbf{break} \Comment{Converged?} \label{alg:rak--converged}
    \EndIf
  \EndFor \label{alg:rak--iters-end}
  \Return{$C$} \label{alg:rak--main-return}
\EndFunction \label{alg:rak--main-end}

\Statex

\Function{lpaMove}{$G, C$} \label{alg:rak--move-begin}
  \State Changed vertices: $\Delta N \gets 0$
  \ForAll{unprocessed $i \in V$ \textbf{in parallel}}
    \State Mark $i$ as processed (prune) \label{alg:rak--prune}
    \State $H_t \gets scanCommunities(\{\}, G, C, i)$ \label{alg:rak--scan}
    \State $\rhd$ Use $H_t$ to choose the most weighted label
    \State $c^* \gets$ Most weighted label to $i$ in $G$ \label{alg:rak--best-community}
    \If{$c^* = C[i]$} \textbf{continue} \label{alg:rak--best-community-same}
    \EndIf
    \State $C[i] \gets c^*$ \textbf{;} $\Delta N \gets \Delta N + 1$ \label{alg:rak--perform-move}
    \State Mark neighbors of $i$ as unprocessed \label{alg:rak--remark}
  \EndFor
\Return{$\Delta N$} \label{alg:rak--move-return}
\EndFunction \label{alg:rak--move-end}

\Statex

\Function{scanCommunities}{$H_t, G, C, i$} \label{alg:rak--scan-begin}
  \ForAll{$(j, w) \in G.edges(i)$}
    \If{$i \neq j$} $H_t[C[j]] \gets H_t[C[j]] + w$
    \EndIf
  \EndFor
  \Return{$H_t$}
\EndFunction \label{alg:rak--scan-end}
\end{algorithmic}
\end{algorithm}

The \texttt{lpaMove()} function (lines \ref{alg:rak--move-begin}-\ref{alg:rak--move-end}) iterates over unprocessed vertices in parallel. For each unprocessed vertex $i$ in the graph $G$, we mark $i$ as processed - vertex pruning (line \ref{alg:rak--prune}), obtain the total edge weight of connected labels in per-thread hashtable $H_t$ with the \texttt{scanCommunities()} function \ref{alg:rak--scan}, and select the most weighted label $c^*$ (line \ref{alg:rak--best-community}). If $c^*$ is not the same as the current label of $i$, we update the label of $i$, increment the count of changed vertices $\Delta N$, and mark the neighbors of $i$ as unprocessed for the next iteration (lines \ref{alg:rak--perform-move}-\ref{alg:rak--remark}). After having processed all vertices, we return the total number of vertices with updated labels $\Delta N$ (line \ref{alg:rak--move-return}). The \texttt{scanCommunities()} (lines \ref{alg:rak--scan-begin}-\ref{alg:rak--scan-end}) iterates over the neighbors of the current vertex $i$, excluding itself, and calculates the total weight of each label in the hashtable $H_t$.

\section{Evaluation}
\label{sec:evaluation}
\subsection{Experimental Setup}
\label{sec:setup}

We employ a server equipped with two Intel Xeon Gold 6226R processors, each featuring $16$ cores running at a clock speed of $2.90$ GHz. Every core is equipped with a $1$ MB L1 cache, a $16$ MB L2 cache, and shares a $22$ MB L3 cache. The system has $376$ GB of memory and runs on CentOS Stream 8. We use GCC 8.5 and OpenMP 4.5. Table \ref{tab:dataset} displays the graphs utilized in our experiments, all sourced from the SuiteSparse Matrix Collection \cite{suite19}.

\begin{table}[hbtp]
  \centering
  \caption{List of $13$ graphs obtained from the SuiteSparse Matrix Collection \cite{suite19} (directed graphs are marked with $*$). Here, $|V|$ is the number of vertices, $|E|$ is the number of edges (after adding reverse edges), $D_{avg}$ is the average degree, and $|\Gamma|$ is the number of communities identified with GVE-LPA.\ignore{In the table, B refers to a billion, M refers to a million and K refers a thousand.}}
  \label{tab:dataset}
  \begin{tabular}{|c||c|c|c|c|}
    \toprule
    \textbf{Graph} &
    \textbf{\textbf{$|V|$}} &
    \textbf{\textbf{$|E|$}} &
    \textbf{\textbf{$D_{avg}$}} &
    \textbf{\textbf{$|\Gamma|$}} \\
    % \textbf{$1 - \Gamma_G$} \\
    \midrule
    \multicolumn{5}{|c|}{\textbf{Web Graphs (LAW)}} \\ \hline
    indochina-2004$^*$ & 7.41M & 341M & 41.0 & 147K \\ \hline  % & \num{4.7e-4} & 2.9 GB
    uk-2002$^*$ & 18.5M & 567M & 16.1 & 383K \\ \hline  % & \num{9.6e-5} & 16 GB
    arabic-2005$^*$ & 22.7M & 1.21B & 28.2 & 213K \\ \hline  % & \num{5.5e-4} & 11 GB
    uk-2005$^*$ & 39.5M & 1.73B & 23.7 & 677K \\ \hline  % & \num{9.6e-5} & 16 GB
    webbase-2001$^*$ & 118M & 1.89B & 8.6 & 6.48M \\ \hline  % & \num{7.3e-7} & 18 GB
    it-2004$^*$ & 41.3M & 2.19B & 27.9 & 611K \\ \hline  % & \num{3.8e-4} & 19 GB
    sk-2005$^*$ & 50.6M & 3.80B & 38.5 & 284K \\ \hline  % & \num{5.8e-4} & 33 GB
    \multicolumn{5}{|c|}{\textbf{Social Networks (SNAP)}} \\ \hline
    com-LiveJournal & 4.00M & 69.4M & 17.4 & 2.19K \\ \hline  % & \num{7.9e-4} & 480 MB
    com-Orkut & 3.07M & 234M & 76.2 & 49 \\ \hline  % & \num{6.7e-2} & 1.7 GB
    \multicolumn{5}{|c|}{\textbf{Road Networks (DIMACS10)}} \\ \hline
    asia\_osm & 12.0M & 25.4M & 2.1 & 278K \\ \hline  % & \num{8.4e-4} & 200 MB
    europe\_osm & 50.9M & 108M & 2.1 & 1.52M \\ \hline  % & \num{6.6e-4} & 910 MB
    \multicolumn{5}{|c|}{\textbf{Protein k-mer Graphs (GenBank)}} \\ \hline
    kmer\_A2a & 171M & 361M & 2.1 & 40.1M \\ \hline  % & \num{9.4e-5} & 3.2 GB
    kmer\_V1r & 214M & 465M & 2.2 & 48.7M \\ \hline  % & \num{3.2e-4} & 4.2 GB
  \bottomrule
  \end{tabular}
\end{table}
% We convert directed graphs (marked with $*$) to undirected by duplicating edges in the reverse direction, and set the weight of each edge to $1$. and $F_{size}$ is size of the \textit{MatrixMarket} file

\begin{figure*}[hbtp]
  \centering
  \subfigure[Runtime in seconds (logarithmic scale) with \textit{FLPA}, \textit{igraph LPA}, \textit{NetworKit LPA}, and \textit{GVE-LPA}]{
    \label{fig:rak-compare--runtime}
    \includegraphics[width=0.98\linewidth]{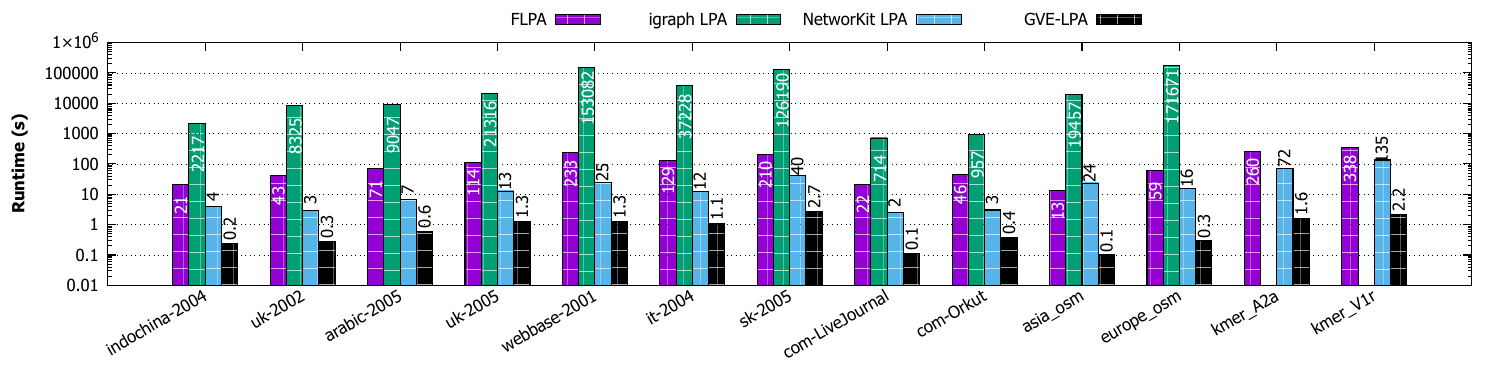}
  } \\[-0ex]
  \subfigure[Speedup (logarithmic scale) of \textit{GVE-LPA} with respect to \textit{FLPA}, \textit{igraph LPA}, \textit{NetworKit LPA}.]{
    \label{fig:rak-compare--speedup}
    \includegraphics[width=0.98\linewidth]{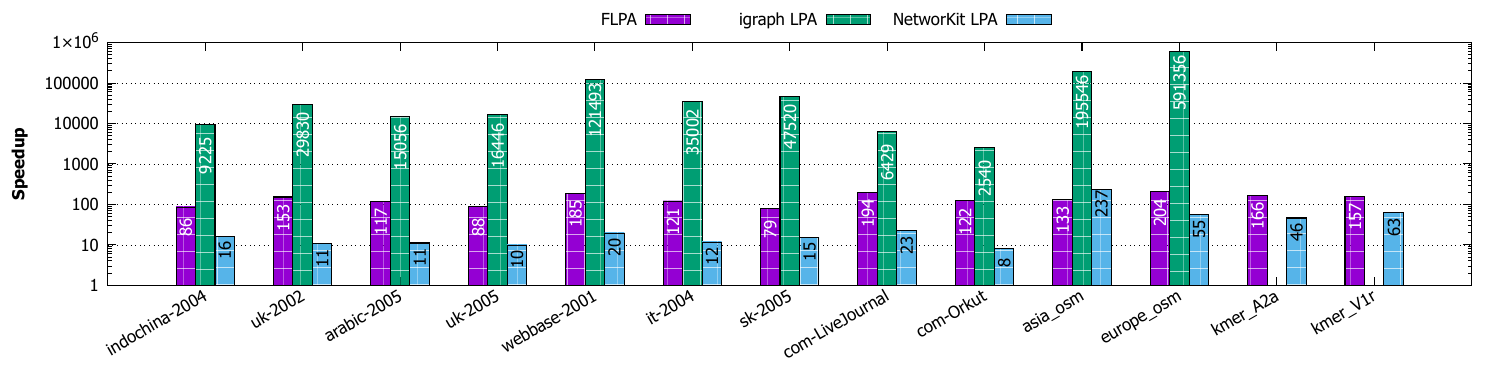}
  } \\[-0ex]
  \subfigure[Modularity of communities obtained with \textit{FLPA}, \textit{igraph LPA}, \textit{NetworKit LPA}, and \textit{GVE-LPA}.]{
    \label{fig:rak-compare--modularity}
    \includegraphics[width=0.98\linewidth]{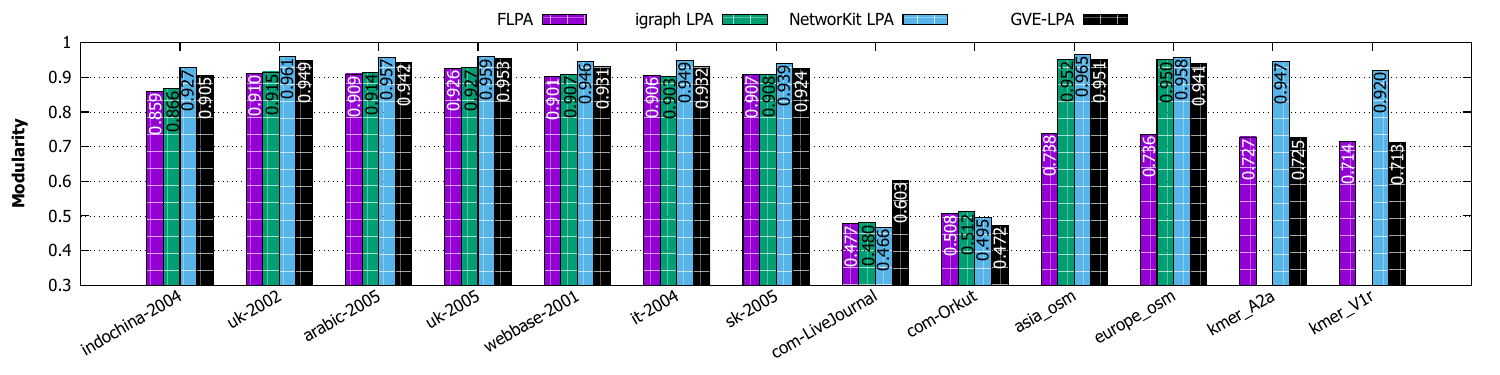}
  } \\[-2ex]
  \caption{Runtime\ignore{in seconds (logarithmic scale)}, speedup, and modularity of communities obtained with \textit{FLPA}, \textit{igraph LPA}, \textit{NetworKit LPA}, and \textit{GVE-LPA} for each graph\ignore{in the dataset}. \textit{FLPA} and \textit{igraph LPA} fail to run on \textit{kmer\_A2a} and \textit{kmer\_V1r} graphs, and thus their results are not shown.}
  \label{fig:rak-compare}
\end{figure*}

\begin{figure*}[hbtp]
  \centering
  \subfigure[Runtime in seconds (logarithmic scale) with \textit{GVE-Louvain} and \textit{GVE-LPA}]{
    \label{fig:louvainrak-compare--runtime}
    \includegraphics[width=0.98\linewidth]{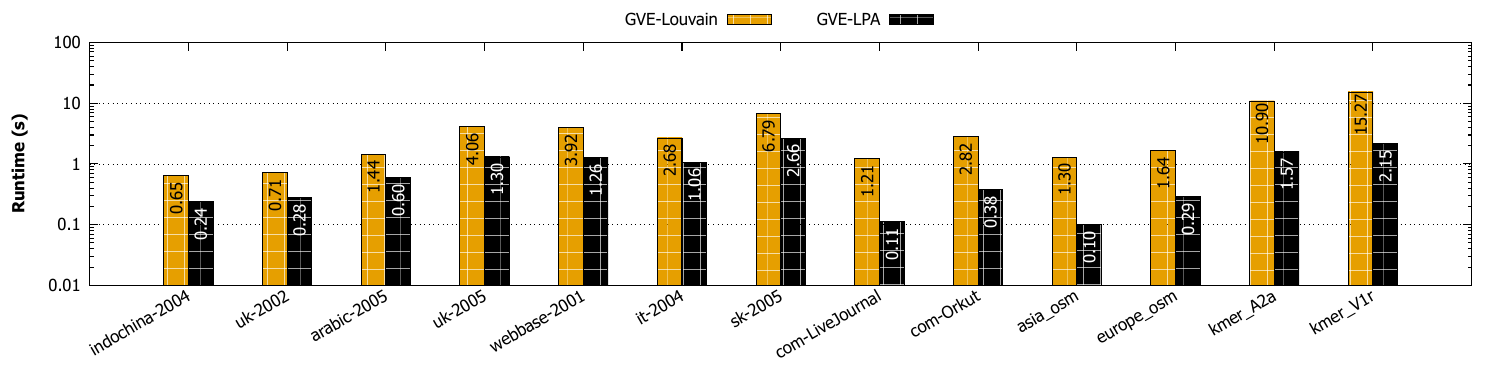}
  } \\[-0ex]
  \subfigure[Speedup of \textit{GVE-LPA} with respect to \textit{GVE-Louvain}.]{
    \label{fig:louvainrak-compare--speedup}
    \includegraphics[width=0.98\linewidth]{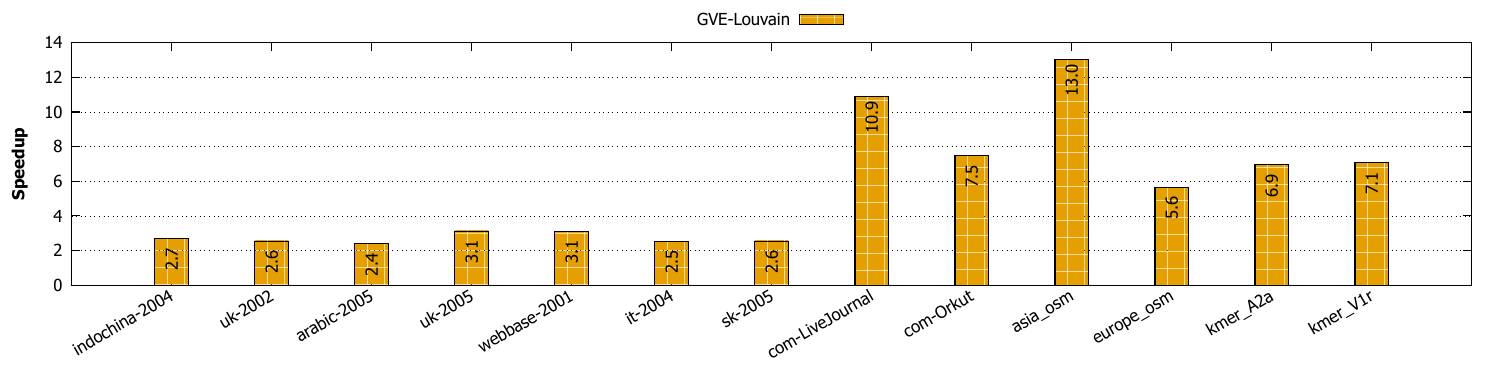}
  } \\[-0ex]
  \subfigure[Modularity of communities obtained with \textit{GVE-Louvain} and \textit{GVE-LPA}.]{
    \label{fig:louvainrak-compare--modularity}
    \includegraphics[width=0.98\linewidth]{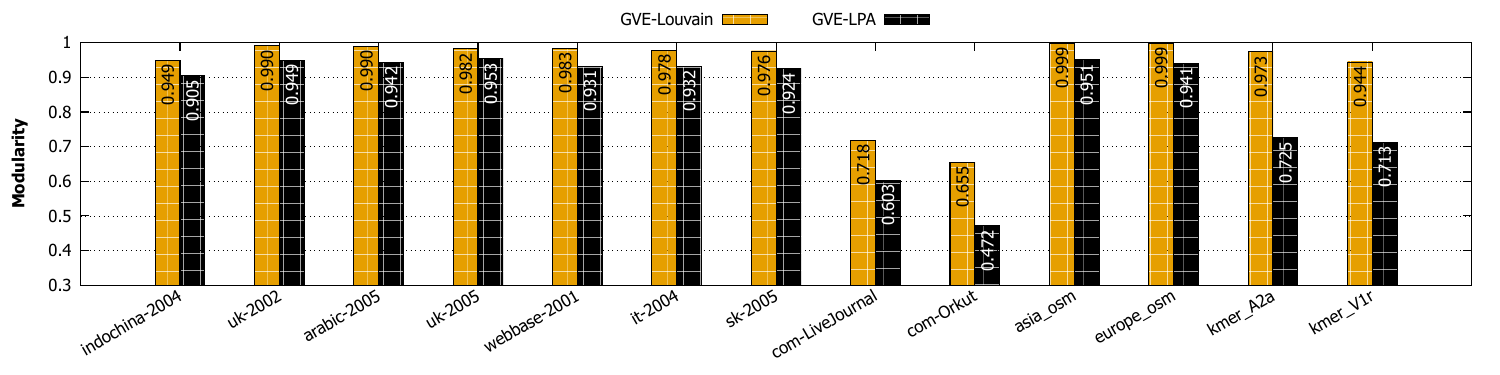}
  } \\[-2ex]
  \caption{Runtime in seconds (logarithmic scale), speedup, and modularity of communities obtained with \textit{GVE-Louvain} \cite{sahu2023gvelouvain} and \textit{GVE-LPA} for each graph in the dataset.}
  \label{fig:louvainrak-compare}
\end{figure*}

\subsection{Comparing Performance of GVE-LPA}

We now compare the performance of GVE-LPA with igraph LPA \cite{csardi2006igraph}, FLPA \cite{traag2023large}, and NetworKit LPA \cite{staudt2016networkit}. igraph LPA and FLPA are both sequential implementations, while NetworKit LPA is parallel. For FLPA, we checkout the suitable branch with modified \texttt{igraph\_community\_label\_propagation()} function, update the label propagation example in C to load the input graph from a file and measure runtime of \texttt{igraph\_community\_label\_propagation} \texttt{()} invoked with \texttt{IGRAPH\_LPA\_FAST} variant, using \texttt{gettimeofday()}. For igraph LPA, we follow a similar process as FLPA, but checkout code from the \texttt{master} branch of igraph. For NetworKit, we employ a Python script to invoke \texttt{PLP} (Parallel Label Propagation), and measure the total time reported with \texttt{getTiming()}. For each graph, we measure the NetworKit LPA runtime five times for averaging. Due to time constraints and long running times of igraph LPA, we ran it only once per graph. Additionally, we record the modularity of communities obtained from each implementation.

Figure \ref{fig:rak-compare--runtime} shows the runtimes of FLPA, igraph LPA, NetworKit LPA, and GVE-LPA for each graph in the dataset, while Figure \ref{fig:rak-compare--speedup} shows the speedup of GVE-LPA with respect to each aforementioned implementation. igraph LPA fails to run on protein k-mer graphs, i.e., \textit{kmer\_A2a} and \textit{kmer\_V1r}, and hence its results on these graphs are not shown. GVE-LPA exhibits an average speedup of $139\times$, $97000\times$, and $40\times$ over FLPA, igraph LPA, and NetworKit LPA respectively. While FLPA and igraph LPA are sequential implementations, GVE-LPA, running on $64$ threads, is more than $64\times$ faster than these implementations. On the \textit{sk-2005} graph, GVE-LPA identifies communities in $2.7$ seconds, achieving a processing rate of $1.4$ billion edges/s. Figure \ref{fig:rak-compare--modularity} shows the modularity of communities obtained by each implementation. On average, GVE-LPA obtains $6.6\%$ / $0.2\%$ higher modularity than FLPA (especially on road networks) and igraph LPA respectively, and $4.1\%$ lower modularity than NetworKit LPA (especially on protein k-mer graphs\ignore{, we plan to address this issue in the future}).

\begin{figure}[hbtp]
  \centering
  \subfigure{
    \label{fig:rak-hardness--all}
    \includegraphics[width=0.98\linewidth]{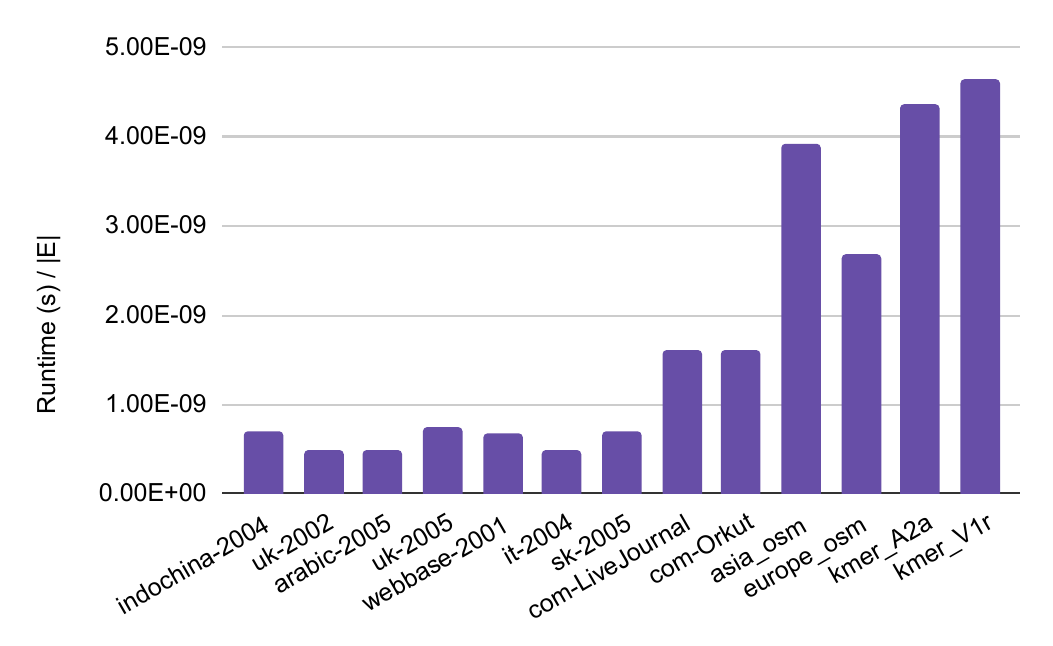}
  } \\[-2ex]
  \caption{Runtime $/ |E|$ factor with \textit{GVE-LPA} for each graph in the dataset.}
  \label{fig:rak-hardness}
\end{figure}

\begin{figure}[hbtp]
  \centering
  \includegraphics[width=0.98\linewidth]{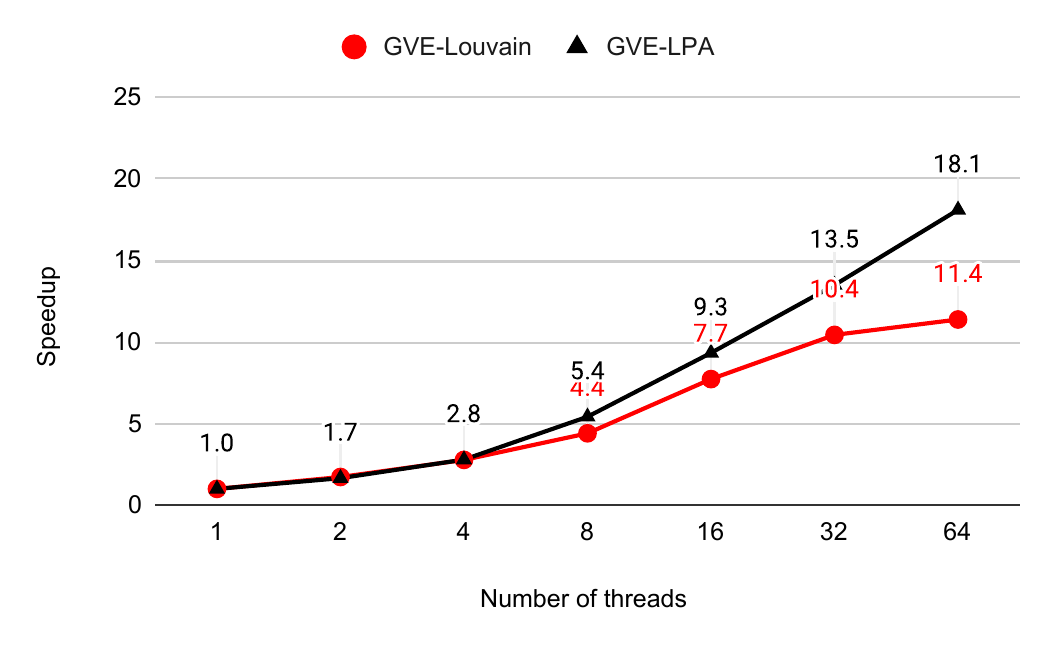} \\[0ex]
  \caption{Average speedup of \textit{GVE-Louvain} and \textit{GVE-LPA} with increasing number of threads (in multiples of 2).}
  \label{fig:rak-ss}
\end{figure}

Next, we proceed to compare the performance of GVE-LPA with GVE-Louvain (our parallel implementation of Louvain algorithm) in Figure \ref{fig:louvainrak-compare}. Figures \ref{fig:louvainrak-compare--runtime}, \ref{fig:louvainrak-compare--speedup}, and \ref{fig:louvainrak-compare--modularity} present the runtimes, speedup (of GVE-LPA with respect to GVE-Louvain), and modularity of GVE-Louvain and GVE-LPA for each graph in the dataset. GVE-LPA offers a mean speedup of $5.4\times$ compared to GVE-Louvain, particularly on social networks, road networks, and protein k-mer graphs, while achieving on average $10.9\%$ lower modularity, especially on social networks and protein k-mer graphs.

In Figure \ref{fig:rak-hardness}, we assess the runtime per edge of GVE-LPA. Results indicate that GVE-LPA exhibits a higher $\text{runtime}/|E|$ factor on graphs with a low average degree (road networks and protein k-mer graphs) and on graphs with a weakly connected clusters (\textit{com-LiveJournal} and \textit{com-Orkut} graphs).

\subsection{Strong Scaling of GVE-LPA}

Finally, we assess the strong scaling performance of GVE-LPA. To this end, we vary the number of threads from $1$ to $64$ in multiples of $2$ for each input graph, and measure the time required for GVE-LPA to identify communities (averaged over five runs). Figure \ref{fig:rak-ss} shows the average strong-scaling of of GVE-LPA, in comparison with GVE-Louvain. With 32 threads, GVE-LPA achieves a speedup of $13.5\times$ compared to single-threaded execution, indicating a performance increase of $1.7\times$ for each doubling of threads. At 64 threads, GVE-LPA is moderately impacted by NUMA effects, and offers a speedup of $18.1\times$. This contrasts with GVE-Louvain, which appears to be significantly affected by NUMA effects.

\section{Conclusion}
\label{sec:conclusion}
In summary, this study focuses on optimizing the Label Propagation Algorithm (LPA), a high-speed community detection algorithm, in the shared memory setting. We consider 6 different optimizations, which significantly improve the performance of the algorithm. Comparative assessments against competitive open-source implementations (FLPA\ignore{\cite{traag2023large}}) and packages (igraph\ignore{\cite{csardi2006igraph}} and NetworKit\ignore{\cite{staudt2016networkit}}) indicate that GVE-LPA outperforms FLPA, igraph LPA, and NetworKit LPA by $139\times$, $97000\times$, and $40\times$ respectively, while identifying communities of $6.6\%$ / $0.2\%$ higher quality\ignore{(modularity)} than FLPA and igraph LPA respectively, and of $4.1\%$ lower quality than NetworKit LPA.

On a web graph with $3.8$ billion edges, GVE-LPA identifies communities in $2.7$ seconds, and thus achieves a processing rate of $1.4$ billion edges/s. GVE-LPA is thus ideal for applications requiring a high-speed clustering algorithm while accepting a compromise in clustering quality --- it is on average $5.4\times$ faster than GVE-Louvain\ignore{\cite{sahu2023gvelouvain}}, and has a strong scaling factor of $1.7\times$ for every doubling of threads, but identifies communities with on average $10.9\%$ lower modularity than GVE-Louvain.\ignore{Future research could explore dynamic algorithms for LPA to accommodate evolving graphs in real-world scenarios. This would allow interactive updation of community memberships of vertices on large graphs.}

%% The acknowledgments section.
\begin{acks}
I would like to thank Prof. Kishore Kothapalli, Prof. Dip Sankar Banerjee, Balavarun Pedapudi, Souvik Karfa, and Vincent Traag for their support.
\end{acks}

%% Bibliography style to be used, and the bibliography file.
\bibliographystyle{ACM-Reference-Format}
\bibliography{main}

\end{document}